\documentstyle[eqsecnum,psfig,aps,epsf]{revtex}

\renewcommand{\baselinestretch}{2.0}
\input{psfig}
\begin{document}
\onecolumn


\vspace*{0.5cm}

\begin{center}
{\large \bf{ Search for second generation leptoquarks in the
    dimuon plus dijet channel of $p\overline{p}$ collisions
    at $\sqrt{s}=1.8$ TeV}} \\
\vspace*{0.5cm}

(July 8, 1998)\\

\end{center}

\font\eightit=cmti8
\def\r#1{\ignorespaces $^{#1}$}
\hfilneg
\begin{sloppypar}
\noindent
F.~Abe,\r {17} H.~Akimoto,\r {39}
A.~Akopian,\r {31} M.~G.~Albrow,\r 7 A.~Amadon,\r 5 S.~R.~Amendolia,\r {27} 
D.~Amidei,\r {20} J.~Antos,\r {33} S.~Aota,\r {37}
G.~Apollinari,\r {31}
% ...
\end{sloppypar}
\vskip .026in
\begin{center}
(CDF Collaboration)
\end{center}

\vskip .026in
\begin{center}
\r 1  {\eightit Argonne National Laboratory, Argonne, Illinois 60439} \\
\r 2  {\eightit Istituto Nazionale di Fisica Nucleare, University of Bologna,
I-40127 Bologna, Italy} \\
\r 3  {\eightit Brandeis University, Waltham, Massachusetts 02254} \\
\r 4  {\eightit University of California at Los Angeles, Los 
Angeles, California  90024} \\  
\r 5  {\eightit University of Chicago, Chicago, Illinois 60637} \\
\r 6  {\eightit Duke University, Durham, North Carolina  27708} \\
\r 7  {\eightit Fermi National Accelerator Laboratory, Batavia, Illinois 
60510} \\
\r 8  {\eightit University of Florida, Gainesville, FL  32611} \\
\r 9  {\eightit Laboratori Nazionali di Frascati, Istituto Nazionale di Fisica
               Nucleare, I-00044 Frascati, Italy} \\
\r {10} {\eightit University of Geneva, CH-1211 Geneva 4, Switzerland} \\
\r {11} {\eightit Harvard University, Cambridge, Massachusetts 02138} \\
\r {12} {\eightit Hiroshima University, Higashi-Hiroshima 724, Japan} \\
\r {13} {\eightit University of Illinois, Urbana, Illinois 61801} \\
\r {14} {\eightit Institute of Particle Physics, McGill University, Montreal 
H3A 2T8, and University of Toronto,\\ Toronto M5S 1A7, Canada} \\
\r {15} {\eightit The Johns Hopkins University, Baltimore, Maryland 21218} \\
\r {16} {\eightit Institut f\"{u}r Experimentelle Kernphysik, 
Universit\"{a}t Karlsruhe, 76128 Karlsruhe, Germany} \\
\r {17} {\eightit National Laboratory for High Energy Physics (KEK), Tsukuba, 
Ibaraki 305, Japan} \\
\r {18} {\eightit Ernest Orlando Lawrence Berkeley National Laboratory, 
Berkeley, California 94720} \\
\r {19} {\eightit Massachusetts Institute of Technology, Cambridge,
Massachusetts  02139} \\   
\r {20} {\eightit University of Michigan, Ann Arbor, Michigan 48109} \\
\r {21} {\eightit Michigan State University, East Lansing, Michigan  48824} \\
\r {22} {\eightit University of New Mexico, Albuquerque, New Mexico 87131} \\
\r {23} {\eightit The Ohio State University, Columbus, OH 43210} \\
\r {24} {\eightit Osaka City University, Osaka 588, Japan} \\
\r {25} {\eightit Universita di Padova, Istituto Nazionale di Fisica 
          Nucleare, Sezione di Padova, I-35131 Padova, Italy} \\
\r {26} {\eightit University of Pennsylvania, Philadelphia, 
        Pennsylvania 19104} \\   
\r {27} {\eightit Istituto Nazionale di Fisica Nucleare, University and Scuola
               Normale Superiore of Pisa, I-56100 Pisa, Italy} \\
\r {28} {\eightit University of Pittsburgh, Pittsburgh, Pennsylvania 15260} \\
\r {29} {\eightit Purdue University, West Lafayette, Indiana 47907} \\
\r {30} {\eightit University of Rochester, Rochester, New York 14627} \\
\r {31} {\eightit Rockefeller University, New York, New York 10021} \\
\r {32} {\eightit Rutgers University, Piscataway, New Jersey 08855} \\
\r {33} {\eightit Academia Sinica, Taipei, Taiwan 11530, Republic of China} \\
\r {34} {\eightit Texas A\&M University, College Station, Texas 77843} \\
\r {35} {\eightit Texas Tech University, Lubbock, Texas 79409} \\
\r {36} {\eightit Istituto Nazionale di Fisica Nucleare, University of Trieste/
Udine, Italy} \\
\r {37} {\eightit University of Tsukuba, Tsukuba, Ibaraki 315, Japan} \\
\r {38} {\eightit Tufts University, Medford, Massachusetts 02155} \\
\r {39} {\eightit Waseda University, Tokyo 169, Japan} \\
\r {40} {\eightit University of Wisconsin, Madison, Wisconsin 53706} \\
\r {41} {\eightit Yale University, New Haven, Connecticut 06520} \\
\end{center}

\begin{abstract}
  \baselineskip 24pt
  
  We report on a search for second generation leptoquarks
  ($\Phi_2$) using a data sample corresponding to an
  integrated luminosity of 110 pb$^{-1}$ collected at the
  Collider Detector at Fermilab.  We present upper limits on
  the production cross section as a function of $\Phi_2$
  mass, assuming that the leptoquarks are produced in pairs
  and decay into a muon and a quark with branching ratio
  $\beta$.  Using a Next-to-Leading order QCD calculation, we
  extract a lower mass limit of $M_{\Phi_2} > 202 (160) $
  GeV$/c^{2}$ at 95\% confidence level for scalar
  leptoquarks with $\beta$=1(0.5).

\end{abstract}



\twocolumn

\narrowtext
\baselineskip 24pt


Leptoquarks are hypothetical bosons which carry both baryon and
lepton quantum numbers and mediate interactions between quarks
and leptons.  They appear in many extensions to the Standard
Model, {\it e.g.} GUT, superstring, horizontal symmetry,
compositeness or technicolor~\cite{othertheory}.
Leptoquarks which combine quarks and leptons of different
generations result in flavor changing neutral currents, which are
known to be highly suppressed~\cite{buckmuller}.  While these
FCNC constraints do not exclude such leptoquarks, they restrict
them to very high masses.  For example, in the Pati-Salam
model~\cite{patisalam}, the masses are expected
in the multi-TeV range, and indirect searches for such
leptoquarks have been made~\cite{flavour}. 
For this search, we assume that the
leptoquarks couple only to leptons and quarks of the same
generation.  This leads to the classification of leptoquarks of
three generations, denoted as $\Phi_i$,~$i=1,2,3$ in this report.

In $p\overline{p}$ collisions, leptoquarks can be pair-produced
by gluon-gluon fusion or $q\bar q$ annihilation~\cite{hewett}.
The contribution to the production rate from direct $\Phi ql$
coupling is suppressed relative to the dominant QCD mechanisms~\cite{hewett}.
The coupling strength to gluons is determined by the color
charges of the particles, and is model-independent in the case of
scalar leptoquarks.  The production of vector leptoquark pairs is
also possible.  However, vector leptoquarks have model-dependent
trilinear and quadratic couplings to the gluon field~\cite{hewett2}.  
In typical cases the production cross section is orders of
magnitude larger than for scalar leptoquarks.  The acceptance for
vector and scalar leptoquark detection is similar, resulting in
much more stringent limits on the vector leptoquark mass.

In this analysis, we report on a direct search for pair produced
second generation scalar leptoquarks. The possible decay channels
are:
\begin{displaymath}
 \begin{array}{ll}
  \Phi_2 \rightarrow q_{2}\mu^{\pm} ,\qquad  
&\mathrm{branching\ ratio}\ \beta \\
  \Phi_2 \rightarrow q_{2}\nu_{\mu} ,\qquad  &\mathrm{branching\ ratio}\ 1 - \beta
 \end{array}
\end{displaymath}
\vspace*{-1.8cm}
\begin{flushright}
(1)
\end{flushright}
where $\beta$ is the branching ratio to charged lepton decay,
and $q_{2}$ is a second generation quark ($c$, $s$).
Our search for $\Phi_2$ production is based on events
having a topology including two muons and at least two jets
\mbox{($\Phi_2\overline{\Phi_2} \rightarrow \mu^{+} \mu^{-} j j$)}. 
The production rate of this decay mode is proportional 
to $\beta^{2}$.


A previous CDF study~\cite{lq2cdf} excluded $M_{\Phi_2} <
131 (96)$~GeV$/c^{2}$, for $\beta=1.0 (0.5)$ using an
integrated luminosity of 19~pb$^{-1}$.  A limit has also
been published by D$\emptyset$~\cite{lq2d0}, which excludes
$M_{\Phi_2}<119 (89) $~GeV/$c^{2}$ for $\beta =1 (0.5)$.
Searches at LEP-1 have excluded leptoquarks with masses
below 45 GeV$/c^{2}$ independent of $\beta$~\cite{lep}.
Here we present a new limit using an integrated luminosity
of 110~pb$^{-1}$ collected during the 1992-93 and 1994-95
Tevatron runs (including the 19~pb$^{-1}$ of the previous CDF
study).  Searches have also been made for first and third
generation leptoquark production at the
Tevatron~\cite{lq13cdf}, LEP~\cite{lep} and
HERA~\cite{heralq}.  The H1 and ZEUS experiments at HERA
have reported the observation of an excess of events at high
$Q^{2}$~\cite{heraq2}. The interpretation of the excess as
the production of a first generation leptoquark has been
ruled out for large $\beta$ by the Tevatron results~\cite{lq13cdf}.
                                                              
The CDF detector is described in full detail
elsewhere~\cite{cdfdetector}.  Only the detector
subsystems that are important in this analysis are mentioned
here.
The momenta of muons are
measured in the Central Tracking Chamber (CTC), a 2.76~m
diameter cylindrical drift chamber.  It is surrounded by a
1.4~T super-conducting solenoidal magnet, covering a
pseudo-rapidity~($\eta$) range up to 1.1, which allows
precision measurements of the transverse momenta ($p_T$) of
charged particles.
Inside the CTC a vertex tracking chamber (VTX) allows event
vertex reconstruction using tracks over the range
$|\eta|<3.25$. Jets are detected by the calorimeters, which
are divided into a central barrel ($|\eta|<1.1$), end plugs
($1.1<|\eta|<2.4$), and forward/backward modules
($2.4<|\eta|<4.2$).  Outside the calorimeters, Central Muon
drift chambers (CMU) in the region $|\eta|<0.6$ provide muon
identification. Outside the CMU lie the Central Muon Upgrade
chambers (CMP), with additional steel between the CMU
and CMP detectors to reduce the background from hadrons in the
muon sample. The region of $0.6<|\eta|<1.0$ is covered
by the Central Muon Extension (CMX) chambers.


We use the PYTHIA Monte Carlo generator~\cite{pythia} with
the {\sc CTEQ4M} parton distribution functions~\cite{pdflib}
and the renormalization and factorization scales defined as
\mbox{$Q^{2} = p^{2}_{T}$}, together with the CDF detector
simulation package, to study the detailed properties of the
signal for $\Phi_2$ masses between 100 and 240~GeV$/c^{2}$.
The signal selection criteria are set according to the
kinematic distributions ({\it e.g.} the $p_{T}$ of the muons
and $E_{T}$ of the jets) of decay products determined by
Monte Carlo studies, optimised to eliminate the background
with a minimal loss of signal events~\cite{sam}.

We select events from several different central single-muon
triggers~\cite{cdfdetector} with \mbox{$p_T$} thresholds
of 9 or 12 GeV/$c$.  From these events an exclusive dimuon
sample is selected by requiring events with two muons
satisfying \mbox{$p_{T} > 30$ GeV$/c$ $(\mu_{1})$} and
\mbox{$p_{T} > 20 $ GeV$/c$ $(\mu_2)$}. We do not require
the two muons to have opposite charge because at very high
$p_{T}$ the charge determination is not reliable.
 One of the muons is
required to have a track from the CTC that matches with a
stub in the fiducial region of the central muon detectors
(within 2 cm for CMU, and 5 cm for CMU/CMP).  The muon
satisfying this criterion is defined as a 'tight' 
muon.
The other muon can be either a tight muon or a 'loose'
muon. A loose muon is defined as a CTC track that deposits less 
than 2 GeV of electromagnetic energy and less than 6 GeV of hadronic 
energy in the calorimeter tower that it traverses. 
To ensure good track quality, the track is required
to traverse at least 75\% of the CTC in the radial
direction, and be matched to an interaction vertex
determined by the VTX to better than 5.0~cm in the $Z$
direction.
Both muons are required to be isolated, defined as \mbox{$I<
  $ 2 GeV}, where $I$ is the sum of transverse energies of
all calorimeter towers (excluding the one traversed by the
muon) within a cone of \mbox{$\Delta R =0.4 $} around the
direction of the muon, where \mbox{$\Delta R =
  \sqrt{(\Delta\eta)^{2} + (\Delta\phi)^{2}}$} and $\phi$ is
the azimuthal angle.

The total dimuon identification efficiency, averaged over
the data sample, lies between 79\% at
$M(\Phi_2)=100$~GeV/$c^{2}$ and 74.5\% at
$M(\Phi_2)=240$~GeV/$c^{2}$, with the dependence on mass
being due to the efficiency of the minimum ionizing
requirement.  The combined average identification and
trigger efficiency is approximately 70\% over the mass range
$100 < M(\Phi_2) < 240$~GeV/$c^{2}$.

From this high-$p_{T}$ dimuon event sample, we require
$\geq$~2 jets with \mbox{$E_{T}^{(1)} > 30 $ GeV} and
\mbox{$E_{T}^{(2)} > 15 $ GeV}, respectively.  Jets are
reconstructed by an algorithm using a fixed cone in
$\eta-\phi$ space. A detailed description of the algorithm
can be found in Ref.~\cite{jet}.  For this analysis a cone
of 0.7 is used.
Both jets are required to be reconstructed in the region
$|\eta|< 2.4$.  Jet energy corrections, due to the
calorimeter non-linearity, energy deposited outside the jet
cone, underlying energy from other interactions, and the
detector geometrical dependence, are applied to determine
the $\mu-jet$ invariant mass.  The $Z^{0}$ and other
resonances such as the $J/\psi$ or $\Upsilon$ are removed by
rejecting events with a dimuon invariant mass in the regions
\mbox{$76 < M_{\mu\mu} < 106 $ GeV$/c^{2}$} and
\mbox{$M_{\mu\mu} < 11 $ GeV$/c^{2}$}.  After applying these
requirements, we are left with a sample of 11 events.

Cosmic rays can fake high-$p_T$ dimuon events; however such
muons take a finite time to traverse the detector, generally
entering from the top of the detector and exiting at the
bottom.  We use the hadronic calorimeter TDC information and
a measurement of the opening angle of the two muons to
reject cosmic ray events. None of the 11 selected events is
identified as a cosmic ray event.

\begin{figure}[h]
\center
\begin{minipage}[t]{7.7cm}
 \epsfxsize=7.5cm
 \mbox{\epsffile{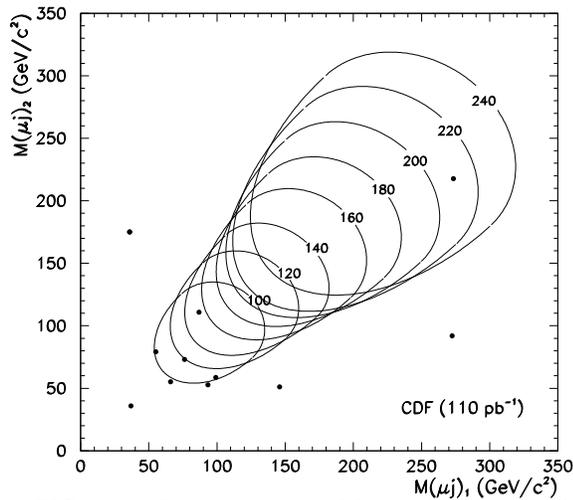}}
\caption{\label{fig_lq2_dist} Invariant mass $M(\mu j)$ 
  distribution for events before applying the mass requirement.  
  A total of 11
  candidate events are displayed on the $M(\mu j)_2$ versus $M(\mu j)_1$
  plane. The oval configuration shows the limit of the mass
  requirement for $\Phi_2$ masses between 100 and 240 GeV/$c^2$.
  $M(\mu j)_1$ is the invariant mass with the higher \mbox{$p_T$} muon,
  while $M(\mu j)_2$ is that with the lower \mbox{$p_T$} muon.}
\end{minipage}
\end{figure}

The numbers of events surviving each selection criterion are
listed in Table~\ref{evtcut}.  A total of 11 events passing
the final selection are shown in Figure~\ref{fig_lq2_dist},
plotted in the muon-jet invariant mass plane \mbox{($M_{\mu
    j}^1 $ v.s.  $M_{\mu j}^2$)}.   From two muons and two
jets, there are two possible muon-jet pairings.  We choose
the combination having the smallest invariant mass
difference to determine the leptoquark mass for possible
candidate events.  The reconstructed leptoquark candidates
of a pair should have equal mass, within the experimental
mass resolution $\sigma_r$.

We therefore search for leptoquark candidates by selecting
events in a $3\sigma_r$ mass resolution region of the $M_{\mu j}^1$ vs.
$M_{\mu j}^2$ plane around any given mass, as shown in
Fig.~\ref{fig_lq2_dist}.  The mass resolution, estimated
from Monte Carlo studies, depends on the event geometry and
the total event energy.  Consequently, it varies with the
leptoquark mass.  For example, the maximum values for the
mass resolutions are
$\sigma_r(\Phi_2=120$~GeV/$c^{2})=21.2$~GeV/$c^{2}$ and
$\sigma_r(\Phi_2=240$~GeV/$c^{2})=46.5$~GeV/$c^{2}$.  The
asymmetric mass resolution (oval-shaped regions shown in
Figure~\ref{fig_lq2_dist}) results primarily from the
detector resolution, but also includes a small probability
of misidentifying the jet when additional jets exist in the
collision, and cases for which a wrong muon-jet pairing
combination is chosen.

The backgrounds for the $\Phi_2$ search include higher order
Drell-Yan processes, heavy flavor decay (from
$\overline{b}b$ or $\overline{t}t$ in the dimuon channel),
$WW$, or \mbox{$Z\rightarrow \tau^{+}\tau^{-}$}.  An
additional background results from $W $ plus multi-jet
events with a fake muon from energetic hadrons penetrating
the shielding to reach the muon chambers, or with a hadron
decay to a muon. These background processes are studied
using relevant Monte Carlo events samples and actual data
samples where possible. The major background is from
Drell-Yan processes (we expect $\sim 12 $ events for 110
pb$^{-1}$) for which the final state includes a muon pair
(\mbox{$Z^{0}/\gamma \rightarrow \mu^{+}\mu^{-}$}) plus two
or more jets from initial or final state radiation.  There
is a small contribution from $\overline{t}t$ (\mbox{$\sim
  1.3 $} events).  Other backgrounds are negligible due to
large muon~$p_T$ and jet~$E_T$ requirements ($\overline{b}b$
and \mbox{$Z\rightarrow \tau^{+} \tau^{-}$}), muon isolation
requirements ($W$ plus jets), and small cross section
($WW$).  The total estimated background is {\mbox{$14\pm
 1.8$}} for an integrated luminosity of 110~pb$^{-1}$, before
applying the 3 $\sigma_{r}$ mass cut.  This mass requirement
reduces the background substantially, since in the
background events the reconstructed muon-jet invariant
masses are not correlated.  For example, we have estimated
the background contribution to be only 0.3 events for
\mbox{M($\Phi_2$) = 200 GeV$/c^{2}$} for the 110~pb$^{-1}$
data sample.

In the final result, we do not apply a background
subtraction procedure, giving the most conservative estimate 
of the cross section limit.  The number of expected events, $N$, is
given by
\begin{displaymath}
  N = {\mathcal L}
    \cdot \beta^{2}
    \cdot \sigma(M_{\Phi_2})
    \cdot \varepsilon_{total},
\end{displaymath}
\vspace*{-2.2cm}
\begin{flushright}
(2)
\end{flushright}
where $\mathcal{L}$ is the total integrated luminosity of
the sample, $\beta$ is the decay branching ratio to the
charged lepton plus quark channel, $\sigma(M_{\Phi 2})$ is
the cross section for a given mass, and $\varepsilon_{tot}$
is the overall efficiency.  We evaluate the factors entering
the overall efficiency as a function of $\Phi_2$ mass using
actual event samples where possible and otherwise simulated
event samples. As shown in Table~\ref{expresult}, it
increases monotonically with M($\Phi_2$), from 9\% at
\mbox{M($\Phi_2$) = 100 GeV$/c^{2}$} to 22\% at
\mbox{M($\Phi_2$) = 240 GeV$/c^{2}$}.

Possible systematic uncertainties of the measured cross section
limit have been studied.  The major source comes from a limited
understanding of the initial and final state gluon radiation.  We
have used Monte Carlo samples with and without gluon radiation to
determine the cross section uncertainty due to this effect.  The
uncertainty decreases as the $\Phi_2$ mass increases, and it is
estimated to be 10\% 
for \mbox{M($\Phi_2$) = 160 GeV$/c^{2}$}.  A
systematic uncertainty also results from the $Q^{2}$ scale and
the structure functions used.  We compute this effect by varying
the $Q^{2}$ scale between 1/4 and 4 of the default value
($Q^{2}=p^{2}_{T}$), and by using other structure functions
(CTEQ2L~\cite{ct2} and MRS(A)~\cite{mrs}).  The jet energy
scaling uncertainty, which results from detector performance
limitations is determined by including a 10\% 
energy uncertainty
in the Monte Carlo reconstruction.  Other sources of uncertainty,
resulting from the detector simulation and the limitation of Monte
Carlo statistics, are relatively small. The uncertainty on 
the luminosity measurement is 7.2\%.  The total systematic uncertainty 
varies with $\Phi_2$ mass, and is computed to be 15\% 
at \mbox{$M_{\Phi 2} = 120 $ GeV$/c^{2}$} and 10\% 
at \mbox{$M_{\Phi 2} = 240 $} GeV$/c^{2}$, 
as listed in Table~\ref{expresult}.

\begin{figure}[h]
\center
\begin{minipage}[t]{7.7cm}
 \epsfxsize=7.5cm
 \mbox{\epsffile{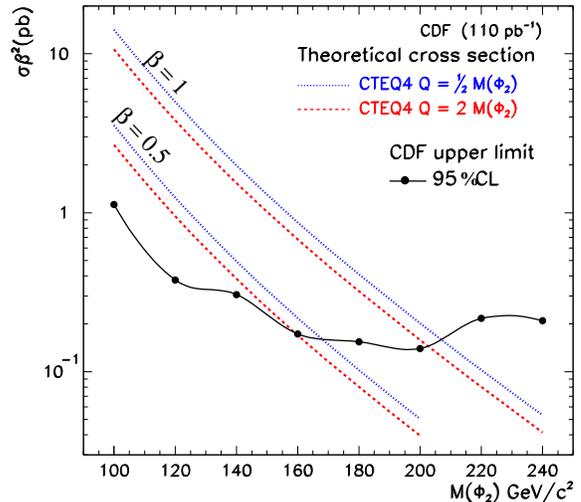}}
  \caption{\label{fig_lq2_limit}                                         
    \small The 95\% 
    C.L. cross section limit for $\Phi_2$ production
    for an integrated luminosity of 110 pb$^{-1}$. 
    The theoretical cross section curves [17] 
    for $\beta$ = 0.5
    and 1.0 are super-imposed.}
  \end{minipage}
\end{figure}

We compute the 95\% confidence level (C.L.) limits on 
\mbox{$\sigma(p\bar p \rightarrow \Phi_2\bar{\Phi}_2)
  \beta^{2}$}, including systematic uncertainties with no
background subtraction, as a function of the leptoquark mass
(see Table~\ref{expresult} and Figure~\ref{fig_lq2_limit}).
The cross section limit for a given $\Phi_2$ mass does not
depend on the coupling $\lambda$, and therefore the mass
limit does not depend on the choice of the theoretical
model but only on $\beta$.  A theoretical 
Next-to-Leading order (NLO) cross section
calculation~\cite{lqxsectheo} is also shown in
Figure~\ref{fig_lq2_limit}, where the band represents the
main uncertainty of the calculation coming from the $Q^{2}$
value.  Comparing the cross section limit to 
this calculation, a limit of
\mbox{$M_{\Phi_2} > 202 (160) $ GeV$/c^{2}$} for
\mbox{$\beta = 1.0 (0.5)$} is derived.
        

We thank the Fermilab staff and the technical staffs of the
participating institutions for their vital contributions.
This work was supported by the U.S. Department of Energy,
the National Science Foundation, the Istituto Nazionale di
Fisica Nucleare (Italy), the Ministry of Science, Culture
and Education of Japan, the Natural Sciences and Engineering
Research Council of Canada, the National Science Council of
the Republic of China, the SchwarzA.~P.~Sloan Foundation and
the A.~von Humboldt-Stiftung, and the Swiss National Science
Foundation.

\renewcommand{\baselinestretch}{1.0}

\newpage


\begin{table}[h]
 \begin{center}
  \begin{tabular}{l|c}
 Type of selection    &   Number of events remaining           \\
 \hline 
 Total number of sample              & 30934 \\
 $1^{st}$ muon selection (tight cut) & 6844  \\
 $2^{nd}$ muon selection (loose cut) & 4153  \\
 2 jet cuts                          & 937   \\
 Jet $E_{T}$ cut                     & 64    \\
 Invariant Mass cut                  & 11    \\
 Cosmic ray cut                      & 11   
 \end{tabular}
 \caption{ \label{evtcut}The number of events surviving each 
   cut, using an integrated luminosity of 110 pb$^{-1}$ 
   from CDF data.  The estimated total
   background from Standard Model sources is $14\pm1.8$ events.}
 \end{center}
\end{table}

\begin{table}[h]
 \begin{center}
  \begin{tabular}{l|cccc}
  $\Phi_2$ mass (GeV$/c^{2}$)      & 120  & 160  & 200  & 240  \\
  \hline
  Total signal detection efficiency,$\varepsilon_{tot}$
                                   & 0.13 & 0.17 & 0.20 & 0.22 \\
  Systematic error on $\varepsilon_{tot}$
                                   & 0.019 & 0.023 & 0.023 & 0.023 \\
  Number of candidate events       & 1    & 0    & 0    & 1    \\
  Estimated background             & 3.8  & 1.1  & 0.3  & 0.1  \\
  $\sigma$ (at 95\% C.L.) in pb    & 0.34 & 0.16 & 0.13 & 0.19 
 \end{tabular}
 \caption{
  \label{expresult}    
  Results for different $\Phi_2$ masses for an integrated
  luminosity of 110 pb$^{-1}$ from CDF data.  No background 
  subtraction is made
  in the cross section evaluation.  The candidate satisfying
  M($\Phi_2$) $>$ 200 GeV/$c^{2}$ was previously
  published [6].}
 \end{center}
\end{table}

\renewcommand{\baselinestretch}{1.0}



\end{document}